\documentclass[12pt]{iopart}

\usepackage{graphicx}   % need for figures
\usepackage{verbatim}   % useful for program listings
\usepackage{color}      % use if color is used in text
\usepackage{hyperref}   % use for hypertext links, including those to external documents and URLs
\usepackage{units}
\usepackage{braket}
\usepackage{siunitx}
\usepackage[noadjust]{cite}
\usepackage{relsize}

\definecolor{grin}{RGB}{19,163,19}

\usepackage[dvipsnames,svgnames,x11names]{xcolor}

\usepackage[final]{changes}

\usepackage[latin1]{inputenc}

\setauthormarkupposition{left}
\setremarkmarkup{(#2)}

\definechangesauthor[color=ForestGreen]{Fred}
\definechangesauthor[color=Orchid]{Valentin}
\definechangesauthor[color=Cerulean]{Vincent}
\definechangesauthor[color=Dandelion]{Laurent}
\definechangesauthor[color=Bittersweet]{Alain}

\newcommand{\tauS}{\tau_\mathrm{s}}
\newcommand{\tauSB}{\tau_\mathrm{s}^\mathrm{Born}}

\newcommand{\tauSscba}{\tau_\mathrm{s}^\mathrm{scba}}
\newcommand{\tauSsf}{\tau_\mathrm{s}^\mathrm{sf}}
\newcommand{\tauScl}{\tau_\mathrm{s}^\mathrm{sf,cl}}

\newcommand{\Ascbacl}{A_\mathrm{scba}^\mathrm{cl}}

\newcommand{\lS}{l_\mathrm{s}}

\newcommand{\VR}{V_\mathrm{R}}

\newcommand{\ki}{\mathbf{k}_i}
\newcommand{\kp}{\mathbf{k}'}
\newcommand{\ks}{\mathbf{k}''}
\newcommand{\kdis}{\mathbf{k}_\mathrm{dis}}
\newcommand{\bfk}{\mathbf{k}}

\newcommand{\sperp}{\sigma}

\newcommand{\SEBornF}{\Sigma_1}
\newcommand{\SEBornS}{\Sigma_2}
\newcommand{\SEscba}{\Sigma_\mathrm{scba}}

\newcommand{\old}[1]{\textcolor{Blue}{\sout{#1}}}
\newcommand{\comm}[1]{\textcolor{Gray}{#1}}

\renewcommand{\old}[1]{}
\renewcommand{\comm}[1]{}
\newcommand{\iogs}{Laboratoire Charles Fabry, Institut d'Optique Graduate School, CNRS, Universit{\'e} Paris-Saclay, 91127 Palaiseau, France}
\newcommand{\MPQ}{ Max-Planck-Institute for Intelligent Systems, Max-Plack-Ring, 4, 72076 T\"ubingen, Germany}
\newcommand{\SAFRAN}{ SAFRAN Sensing Solutions, Safran Tech, Rue des Jeunes Bois, Ch\^{a}teaufort CS 80112, 78772 Magny-les-Hameaux, France}
\newcommand{\Lih}{Zhejiang Institute of Modern Physics, Zhejiang University, Hangzhou 310027, P. R. China}
\newcommand{\KIP}{Heidelberg University, Kirchhoff-Institut f\"{u}r Physik, Im Neuenheimer Feld 227, 69120 Heidelberg, Germany}
\newcommand{\CPHT}{CPHT, CNRS, Ecole Polytechnique, Institut Polytechnique de Paris, Route de Saclay, 91128 Palaiseau, France}
\begin{document}

\title[Elastic scattering time in the strong scattering regime]{Ultracold atoms in disordered potentials: elastic scattering time in the strong scattering regime }
\author{Adrien Signoles$^1$, Baptiste Lecoutre$^1$, J{\'e}r{\'e}mie Richard$^1$, Lih-King Lim$^{2,1}$, Vincent Denechaud$^{1,3}$, Valentin V. Volchkov$^{1,4}$, Vasiliki Angelopoulou$^1$, Fred Jendrzejewski$^{1,5}$, Alain Aspect$^1$, Laurent Sanchez-Palencia$^6$, Vincent Josse$^1$}

\address{$^1$ \iogs}
\address{$^2$ \Lih}
\address{$^3$ \SAFRAN}
\address{$^4$ \MPQ}
\address{$^5$ \KIP}
\address{$^6$ \CPHT}

\ead{vincent.josse@institutoptique.fr}

\begin{abstract}

We study the elastic scattering time $\tauS$ of ultracold atoms propagating in optical disordered potentials in the strong scattering regime, going beyond the recent work of J. Richard \emph{et al.} \textit{Phys. Rev. Lett.} \textbf{122} 100403 (2019). There, we identified the crossover between the weak and the strong scattering regimes by comparing direct measurements and numerical simulations to the first order Born approximation. Here we focus specifically on the strong scattering regime, where the first order Born approximation is not valid anymore and the scattering time is strongly influenced by the nature of the disorder. To interpret our observations, we connect the scattering time $\tauS$ to the profiles of the spectral functions that we estimate using higher order Born perturbation theory or self-consistent Born approximation. The comparison reveals that self-consistent methods are well suited to describe $\tauS$ for Gaussian-distributed disorder, but fails for laser speckle disorder. For the latter, we show that the peculiar profiles of the spectral functions, as measured independently in V. Volchkov \emph{et al.} \textit{Phys. Rev. Lett.} \textbf{120}, 060404 (2018), must be taken into account. Altogether our study characterizes the validity range of usual theoretical methods to predict the elastic scattering time of matter waves, which is essential for future close comparison between theory and experiments, for instance regarding the ongoing studies on Anderson localization.

\end{abstract}

%\submitto{\NJP}
 
\maketitle

\section{Introduction}

Ultracold atoms propagating in disordered potentials offer controllable platforms to study a large variety of quantum transport phenomena~\cite{lsp2010,lewenstein2012}, from the celebrated Anderson localization at the single particle level~\cite{Billy2008,Roati2008,Jendrzejewski2012,Semeghini2015}, to the study of superfluid to insulator transitions~\cite{Derrico2014,Brantut2012,Krinner2013} or the concept of many-body localization~\cite{Schreiber2015,Nandkishore2015,Choi2016} for interacting atoms. One of the major interest of these systems is the ability to confront directly experiments and  theory for a wide range of parameters. In this context, the precise knowledge of the elastic scattering time $\tauS$, which corresponds to the mean time between two scattering events, is essential. 
This fundamental time scale is indeed at the heart of our basic understanding of wave propagation in disordered media, and it is used by theoreticians as an elementary building block in order to elaborate quantitative descriptions of these complex systems~\cite{Kuhn2005,Kuhn2007,Hartung2008,Skipetrov2008,Yedjour2010,Cherroret2012,Piraud2012,Shapiro2012,Plisson2013,Piraud2013,Piraud2014}. 

However, while the elastic scattering time can be predicted with a rather good confidence in the weak scattering regime using perturbative approaches, much less is known in the strong scattering regime~\cite{Rammer2004,Akkermans2007,Lagendijk1996,Skipetrov2018}. One enters this regime, which is the one of interest for Anderson localization, when the mean free path becomes smaller than the (de Broglie) wavelength, i.e., when passing the well-known Ioffe-Regel like criterion $k \lS \sim 1$ ($k$: wave number, $\lS=v\tauS$: mean free path, $v$ being the group velocity). Despite a large amount of work, either with electronic waves \cite{Niederer1974,Ando1982,Bockelmann1990,Monteverde2010} or classical waves~\cite{Shapiro1986,Page1996,Sebbah2002,Emiliani2003,AnacheMenier2009,Jacques2012,Sevrain2013,Obermann2014,Hildebrand2014,Savo2017,Martin2016}, a complete description of $\tauS$ relying on a close comparison between theory and experiments is still lacking. 

In a recent paper~\cite{Richard2019}, we made an important step into that direction. There, the elastic scattering time of ultracold atoms in laser speckle disordered potential was directly measured over a very broad range of experimental parameters, and found to be in excellent agreement with numerical simulations. By comparing the deviations of $\tauS$ to first order Born calculations~\cite{Piraud2013,Piraud2014}, we have identified the crossover between the weak and the strong scattering regime, revealing that its location is strongly influenced by the disorder statistics. 
This was done by using both attractive or repulsive laser speckle disordered potentials, 
whose amplitude probability distributions follow exponential laws,
and by complementing our study by a numerical investigation of a Gaussian-distributed random potential, as usually considered in condensed matter~\cite{Rammer2004,Akkermans2007}.

Here we focus on the description of the mean scattering time in the strong scattering regime, where the first order Born approximation is not valid anymore. To do so, we relate our measurements of $\tauS$ to the width of the spectral functions. These functions give the energy-momentum relation for one particle excitation~\cite{Bruus2004}. They are estimated for our specific system via two different approaches: either by extending the perturbative Born expansion to higher order terms~\cite{Akkermans2007,Kuhn2007,Lugan2009} or by the use of the self consistent Born approximation (SCBA)~\cite{Skipetrov2008,Yedjour2010}. While we find that the perturbative approach allows us to extend the quantitative prediction of $\tauS$ only in a limited range, a first important result is the striking agreement obtained between the SCBA predictions and the mean scattering time for the Gaussian-distributed disorder case. 
However this method cannot cope with the specific statistics of the laser speckle potentials, for which large deviations are observed. To get further insight, we show in a second step that these deviations can be traced to the peculiar behavior of the spectral functions for such disordered potentials~\cite{Trappe2015,Prat2016}. Indeed, we recover full consistency between our measurements of $\tauS$ and the width of the spectral functions when considering the real profiles that have been measured independently using an radio-frequency spectroscopic method, see Ref.~\cite{Volchkov2018}. 

The manuscript is organized as follows. In section~\ref{sec:Born}, we review the measurements of the elastic scattering time and the comparison with the 1st order Born approximation as presented in Richard \emph{et al.}~\cite{Richard2019}. Section~\ref{sec:spectral_functions} provides the adequate framework, based on the direct connection between time properties and spectral functions, to further describe elastic scattering time beyond the first order Born approximation. Finally, we link in section~\ref{sec:red_blue_sf} our observations of elastic scattering time with experimentally obtained profiles of the spectral functions, both for attractive and repulsive laser speckle disorders.

\section{Elastic scattering time along the crossover from weak to strong scattering}
\label{sec:Born}

Using ultracold atoms propagating in optical disordered potentials, we experimentally and numerically determined in Ref.~\cite{Richard2019} the scattering time $\tauS$ of a matter wave launched in a disordered potential $V(\mathbf{r})$ with a well-defined momentum $\ki$. By exploring a broad range of microscopic parameters, we collected an extensive set of data that we use all along this study as a support to explore the behavior of $\tauS$ in the strong scattering regime. This section reviews the main results of Ref.~\cite{Richard2019}, especially the comparison with the first order Born predictions, providing all the details relevant to the remainder of this work.

\subsection{Lifetime of excitation in disorder: the first order Born approximation}
\label{subsec:FGR}

In the weak scattering regime, the propagation of a wave can be described as a succession of independent scattering events that are separated on average by a time $\tauS$, and between which the wave freely propagates. This approximation is known as the first order Born approximation, since it can be obtained by restricting the Born pertubative series to its first order (see~\sref{sec:spectral_functions} for more details)~\cite{Rammer2004,Akkermans2007}. 
In this simple picture, each event results in a transfer from the initial momentum state $\ket{\ki}$ toward a continuum of final momentum states $\ket{\kp}$, with $|\kp| = |\ki|$, see~\fref{fig:decay}(a). The elastic scattering time $\tauS$ can be then interpreted as the lifetime of the initial state $\ket{\ki}$, this time being inversely proportional to the transfer rate to the continuum. The population $\tilde{n}_i(t)$ of $\ket{\ki}$ is thus expected to decay exponentially with time $t$, with a characteristic time $\tauS$:

\begin{equation}
\label{eq:exponential_decay}
\tilde{n}_i(t)=\tilde{n}_i(0)\,e^{-t/\tauS} \,.
\end{equation} 

The scattering time $\tauS$ can be calculated using Fermi golden rule. The coupling rate $\overline{|\bra{\ki}V\ket{\kp}|^2}$ (where $\overline{\cdots}$ refers to disorder averaging) to each state $\ket{\kp}$ is given by the spatial frequency distribution of the disorder $\tilde{C}(\ki-\kp)$. Here, $\tilde{C}$ refers to the Fourier transform of the two-point correlation function $C(\Delta \mathbf{r})=\overline{ \delta V(\mathbf{r}) \delta V(\mathbf{r} +\Delta \mathbf{r})}$, with $\delta V(\mathbf{r})= V(\mathbf{r}) - \overline{V(\mathbf{r})}$ the fluctuations of the disordered potential. It leads to an estimate of the elastic scattering time in the Born approximation
\begin{equation}
\label{eq:fermi_golden_rule}
\frac{\hbar}{\tauSB}= 2\pi \sum_{\kp}\tilde{C}(\ki-\kp)\  \delta( \mathcal{E}_{k_i}-\mathcal{E}_{k'} )\,,
\end{equation}
where $\mathcal{E}_k=\hbar^2k^2 /2m$ is the free-state energy, with $m$ the atomic mass and $\hbar=h/2\pi$ the reduced Plank constant. Hence $\tauSB$ depends only on the spatial correlations of the potential given by $C(\Delta \mathbf{r})$. Since its amplitude is proportional to $|\VR|^2$, with $|\VR|$ the rms value of the disorder potential, an important feature of \eref{eq:fermi_golden_rule} is the simple $1/|\VR|^{2}$ scaling. As a direct consequence, the Born prediction $\tauSB$ is not sensitive to the specific form of the amplitude probability distribution $P(V)$.

\subsection{Measurements of the elastic scattering time}

\begin{figure}[t]
	\includegraphics[width=0.98\textwidth]{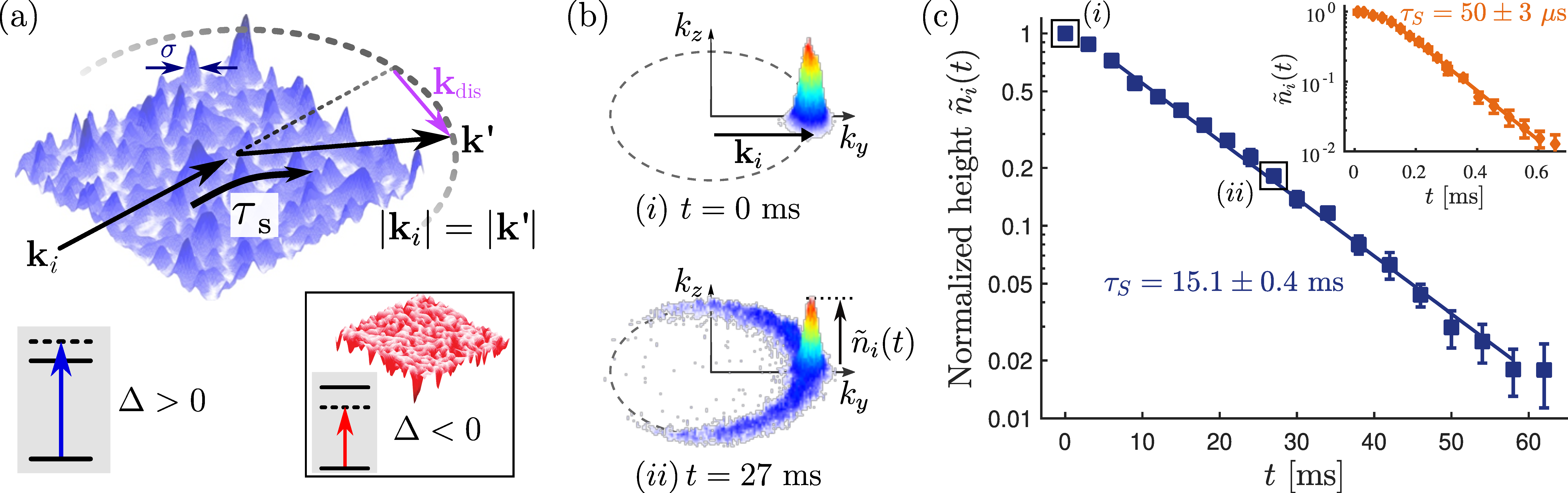}
	\caption{\label{fig:decay} \textbf{Elastic scattering time and Born approximation.}
	(a) Illustration of a scattering event in the Born approximation. An initial momentum state $\ket{\ki}$ is scattered by the potential towards a momentum state $\ket{\kp}$ after a mean time $\tauS$. The repulsive potential, shown in blue, is generated by a laser blue-detuned from an atomic transition. An attractive potential can be generated with a red-detuned laser (see inset).
	(b) Measured momentum distribution $n(\bfk,t)$, for $k_i=1.62\sigma^{-1}$ and $|\VR|/h=\SI{72}{\hertz}$. At time $t=0$ we see the initial momentum distribution of the state $\ket{\ki}$. After a time evolution $t=27$ ms, the wave has been partially scattered, resulting in a reduced peak at $\bfk=\ki$ on top of a ring of radius $k=k_i$. The height of the peak, normalized by its value at $t=0$, gives the population of the initial state $\tilde{n}_i(t)$.
	(c)~Evolution with time $t$ of the population $\tilde{n}_i(t)$ (dots) for $k_i=1.62\sigma^{-1}$ and $\VR/h=\SI{72}{\hertz}$ (inset: $\VR/h=\SI{1.30}{\kilo\hertz}$). The solid line is an exponential fit from which we extract $\tauS$. 
	}
\end{figure}

As discussed in Ref.~\cite{Richard2019}, we experimentally measure $\tauS$ by monitoring the decay of the population in the initial momentum state $\ket{\ki}$ given by~\eref{eq:exponential_decay}. The experimental setup relies on an ultracold, non-interacting Bose-Einstein condensate of $^{87}$Rb that expands in a quasi-2D laser speckle field~\cite{Clement2006,Goodman2007}. We prepare the atoms with an initial momentum $\bfk_i$ along the $y$ direction, the norm $k_i$ ranging from 1 to $20\,\rm{\mu m}^{-1}$, by pulsing an external magnetic gradient for a tunable duration. The laser wavelength for the speckle can be either red- or blue-detuned with respect to the atomic transition, yielding attractive or repulsive disordered potentials (see~\fref{fig:decay}(a)). They exhibit inverted amplitude probability distributions that follow the asymmetrical exponential laws $P_\mathrm{sp}(V)=| \VR |^{-1} e^{-V / \VR} \cdot \Theta (V/ \VR ) $, with  $\Theta$ the step function. While the averaged amplitude is given by $\VR$ (negative for attractive and positive for repulsive laser speckle), the disorder strength is characterized by the rms disorder amplitude $|\VR|$. It can be tuned from $|\VR|/h=\SI{39}{Hz}$ to $\SI{3.88}{kHz}$ by varying the laser power and detuning. The laser speckle is shine on the atoms along the $x$ axis, resulting in a very elongated speckle pattern in this direction. This yields to a quasi-2D disorder geometry in the $(y-z)$ plane, whose transverse two-point correlation function $C(\Delta \mathbf{r})$ is found to have a Gaussian shape of size $\sigma=0.50(1)$~$\mu$m (1/e radius), see Supplemental Material of Ref.~\cite{Richard2019}.

To extract $\tauS$, we record the momentum distribution $n(\mathbf{k},t)$ at different time $t$ by performing fluorescence imaging after a long time-of-flight (see~\fref{fig:decay}(b)). The overall momentum resolution is $\Delta k=\SI{0.2}{\mu\meter}^{-1}$, limited by the finite temperature of the initial state and imaging resolution. From those images we monitor the decay of the population in the initial momentum state $\tilde{n}_i(t)$ (see~\fref{fig:decay}(c))~\cite{Richard2019}. For weak scattering, we observe an exponential decay over typically two orders of magnitude, which we fit with~\eref{eq:exponential_decay} to extract the value of $\tauS$. Although the exponential decay is not expected to persist beyond the Born approximation (see e.g.,~\cite{Skipetrov2018}), at strong scattering we do not observe significant deviations from such a decay within the experimental error bars (see inset in~\fref{fig:decay}(c)). The extraction procedure is thus kept the same over the whole range of parameters.

%%%%%%% Numerics %%%%%%%
The experimentally measured $\tauS$ are plotted in figure~\ref{fig:born} for both attractive (left panel) and repulsive (middle panel) laser speckle disorder. The broad range of parameters $k_i$ and $\VR$ we explore allows us to observe variations of $\tauS$ over more than three orders of magnitude. We compare the measurements with numerical simulations, performed by propagating in time a wave packet of initial momentum $\ki$ in a purely 2D disordered potential (solid lines). The agreement is in general very good and confirms the excellent control over the experimental parameters. It also highlights the quasi-2D nature of our geometry. For simplicity, we thus only compare in the following our measurements to purely 2D theoretical predictions~\footnote{As detailed in the supplemental material of Ref.~\cite{Richard2019}, no significant deviations were found between 3D and 2D calculations for our configuration.  }.

Numerically, we have also explored disordered potential with Gaussian amplitude probability distribution $P_g(V)=(\sqrt{2\pi}\VR)^{-1}e^{-V^2/(2\VR^2)}$ (right panel in~\fref{fig:born}). The two-point correlation function $C(\Delta \mathbf{r})$ is chosen to have a Gaussian shape of size $\sigma$, i.e., to be the same as for the laser speckle~\cite{Richard2019}. It is indeed of primordial interest to further explore the role of disorder statistics, in particular because Gaussian-distributed disorder is the model usually considered in condensed matter~\cite{Rammer2004,Akkermans2007}. Such potential could be also implemented in our experiment using spatial light modulators~(see e.g.~\cite{Choi2016}).

\subsection{Comparison to Born prediction}

\begin{figure}[t]
	\includegraphics[width=\textwidth]{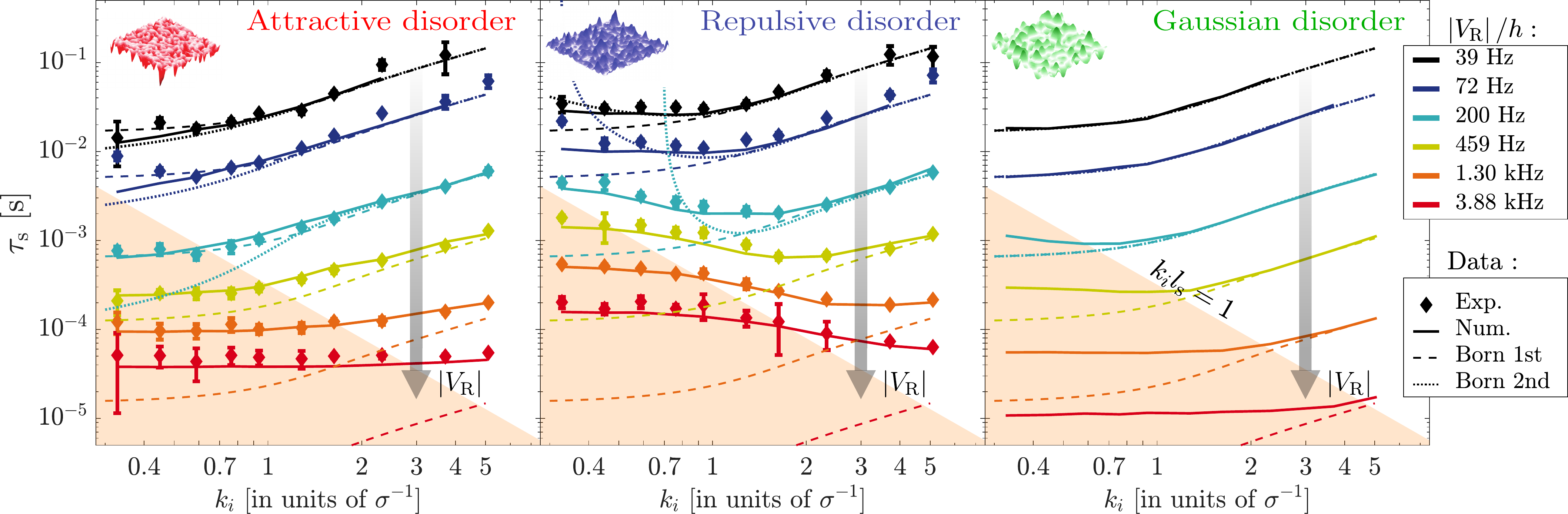}
	\caption{\label{fig:born} \textbf{Experimental and numerical determination of $\tauS$.} 
	Experimental measurements (dots) and numerical simulations (solid lines) of $\tauS$ as a function of the initial momentum $k_i$ for different values of the disorder strength $|\VR|$, in the cases of attractive disorder (left panel), repulsive disorder (central panel) or Gaussian disorder (right panel). The first order Born approximation~\eref{eq:fermi_golden_rule} appears in dashed lines, while the second order Born approximation is shown in dotted lines (only for the three first disorder strengths). The initial momenta are shown in units of the characteristic frequency $\sperp^{-1}$ of the disorder. 
	The shaded area indicates the strong scattering regime $k_i\lS < 1$.  	}
\end{figure}

We compare the experimental and numerical data to the prediction of the first order Born approximation $\tauSB$ given by~\eref{eq:fermi_golden_rule} (dashed lines in \fref{fig:born}). As already mentioned, the prediction is identical for the three types of disorder since they have the same two-point correlation function $C(\Delta\mathbf{r})$. Note that when changing the disorder strength $|\VR|$, the curves in the vertical logarithmic scale are simply shifted down according to the scaling $\tauSB \propto 1/|\VR|^{2}$.

As expected, the agreement is very good  for all three types of disorder in weak scattering regime $k_i\lS \gg 1$, corresponding to low disorder strength $|\VR|$ and large initial momentum $k_i$. For increasing scattering strength, distinct behaviors are observed between Gaussian-distributed and laser speckle disorders. For Gaussian-distributed disorder, the good agreement persists up to $k_i\lS \sim 1$ (indicated by the limit of the shadded area in~\fref{fig:born}). It validates the latter as an accurate criterion to estimate the position of the crossover between weak and strong scattering regimes~\cite{Richard2019}. For laser speckle disorders, however, the Born approximation fails at much lower scattering strength. A quantitative analysis of the deviations performed in Richard \emph{et al.} shows that the position of the crossover is shifted up to $k_i\lS \sim 40$~\cite{Richard2019}. In addition, we note substantial differences between attractive and repulsive laser speckle disorder. The latter, commonly used in the experimental studies of Anderson localization~\cite{Billy2008,Jendrzejewski2012,Semeghini2015}, leads to much larger deviations from the Born prediction in the strong scattering regime.

The emergence of differences between the three types of potential indicates the break down of the first order Born approximation, revealing that the elastic scattering time becomes sensitive to higher-order correlation functions~\cite{Akkermans2007}. To push further theoretical investigation of the elastic scattering time, we develop the connection between time evolution of the system and spectral properties, which requires introducing the concept of spectral functions~\cite{Bruus2004}.

\section{Scattering time and spectral functions of matter waves in disordered potentials}
\label{sec:spectral_functions}

The spectral function $A(\mathcal{E},\ki)$ gives the probability distribution for an excitation of momentum $\ki$ to have a certain energy $\mathcal{E}$, thereby generalizing the concept of dispersion relation. It is for instance used to describe quasi-particles in many-body physics~\cite{Bruus2004,Damascelli2004,Stewart2008,Clement2009,Gaebler2010,Ernst2010,Feld2011} or in disordered systems~\cite{Kuhn2007,Skipetrov2008,Yedjour2010,Trappe2015,Prat2016,Volchkov2018}. Of particular interest for the latter is the width of the spectral function, which is related to the time scale of the scattering processes.

In order to get a intuitive understanding of this fundamental link, it is worth to consider once again the  weak disorder picture. In the absence of disorder, an excitation of well-defined momentum $\ket{\ki}$ is an eigenstate of the Hamiltonian with infinite lifetime: it has a well-defined energy and the spectral function is a Dirac distribution centered on the kinetic energy $\mathcal{E}_{k_i}$. When propagating into a weak disordered potential, the excitation $\ket{\ki}$ is no longer an eigenstate: it acquires a finite lifetime $\tauSB$, given by the Fermi golden rule \eref{eq:fermi_golden_rule}, which translates in energy space into a Lorentzian spectral function $A(\mathcal{E},\ki)$ of finite width $\hbar/\tauSB$. This link between energy width and time scale remains formally relevant even for strong scattering regimes. Provided that the spectral function has no apparent substructures~\footnote{In the case of multiple substructures, the scattering processes are characterized by various timescales, as seen in~\ref{sec:red_blue_sf} for repulsive laser speckle disorder.}, it is indeed always possible to define a characteristic time
\begin{equation}
	\tauSsf=\hbar/\Delta \mathcal{E}
	\label{eq:tauSsf}
\end{equation}
based on the full-width at half-maximum (FWHM) $\Delta \mathcal{E}$ of $A(\mathcal{E},\ki)$, regardless of its exact profile. In the following, our approach consists in confronting different theoretical estimates of this timescale to the scattering time $\tauS$ that we extracted from the decay of time evolution (see~\sref{sec:Born}).

To do so, we first present (\sref{subsec:SE_formalism}) some basic features of the spectral functions and we discuss the expected profiles associated to the various scattering regimes. We then investigate how relevant are perturbative treatments (sections \ref{subsec:SE_Born} and~\ref{subsec:SE_Born2}) and self-consistent Born theory (\sref{subsec:SE_scba}) in describing, within this framework, the scattering time $\tauS$ beyond the weak scattering regime.

\subsection{Generalities about spectral functions}
\label{subsec:SE_formalism}

For disordered systems, the spectral function is generally defined from the averaged Green function $\bar{G}$ as
\begin{equation}
	A(\mathcal{E},\ki) = -\frac{1}{\pi}\text{Im}[\bar{G}(\mathcal{E}, \ki)]\, .
	\label{eq:spec_func_def}
\end{equation}
The calculation of the spectral functions is thus directly related to the one of $\bar{G}(\mathcal{E}, \ki)$ and we briefly review below the main steps of the derivation. 

In the absence of disorder, the system is characterized by the free Green function $G_0(\mathcal{E},\ki)=(\mathcal{E} - \mathcal{E}_{k_i} + i0^+)^{-1}$, and the spectral function is indeed a Dirac function with infinitely small width. 
When taking into account the presence of a disordered potential, the averaged Green function $\bar{G}$ cannot be easily determined. In a general way, the effect of the disordered potential on $\bar{G}$ is encoded into a complex self-energy $\Sigma$, defined by the relation
\begin{center}
	\begin{equation}
	\bar{G}(\mathcal{E},\ki)= \left( \mathcal{E} - \mathcal{E}_{k_i} - \overline{V} - \Sigma(\mathcal{E},\ki)\right)^{-1},
	\label{eq:av_green_func}
	\end{equation}
\end{center}
where we have explicitly isolated the energy shift $\overline{V}$ associated to the mean energy of the potential, such that $\Sigma$ is only associated to the disorder fluctuations~\cite{Yedjour2010}. Determining the self energy is a complex task and it is in general not possible to have an exact expression. Various theoretical approaches render possible its estimate in certain regimes, such as perturbative treatments using a  Born expansion~\cite{Kuhn2007,Lugan2009}, self-consistent approximations~\cite{Skipetrov2008,Yedjour2010} or semi-classical considerations~\cite{Trappe2015,Prat2016}.

\begin{figure}[t]
	\centering
	\includegraphics[width=0.98\textwidth]{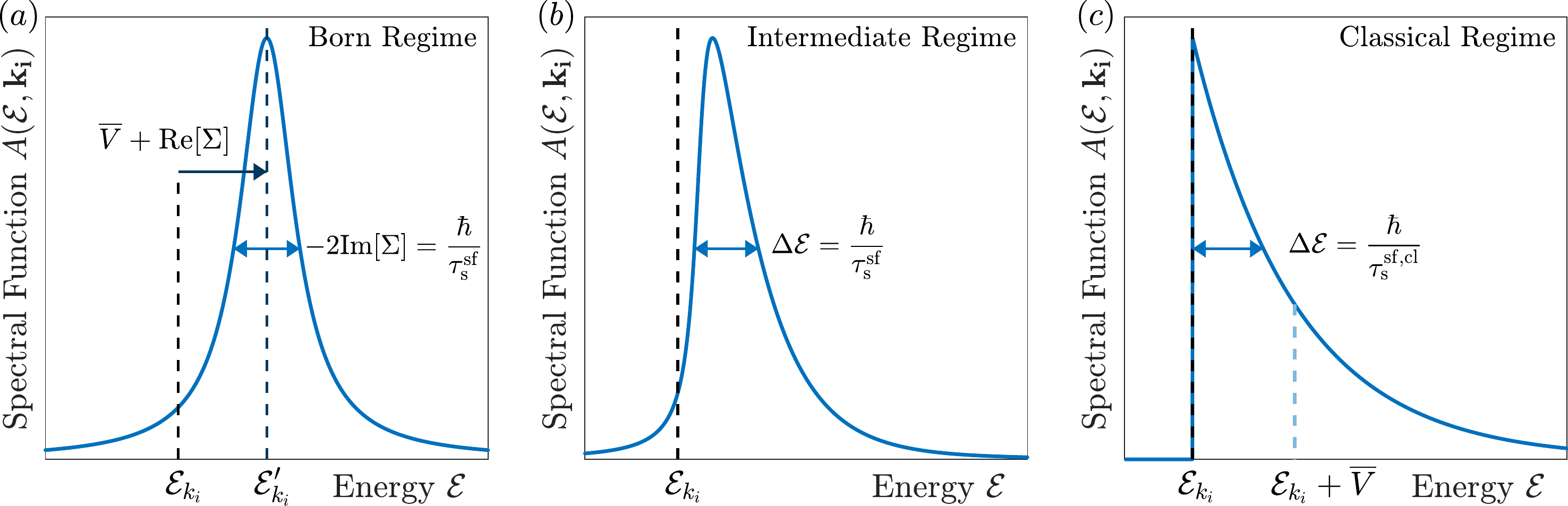}
	\caption{\label{fig:theory_sf} \textbf{Illustration of the profiles of the spectral function in different regimes in the case of repulsive laser speckle disorder.}
	(a) In the Born regime, $A$ is a Lorentzian function with FWHM $\Delta\mathcal{E}=-2\text{Im}[\Sigma]$, inversely proportional to the scattering time $\tauSsf$. (b) In the intermediate regime, no general predictions can be done about the profile of the spectral functions, but an effective scattering time $\tauSsf$ can be defined from the FWHM $\Delta\mathcal{E}$. (c) In the classical limit of strong disorder strength, $A$ approaches the probability distribution of the potential $P(V)$. The effective scattering time $\tauSsf$ converges towards the classical limit $\tauScl$.
	}
\end{figure}

Without going further on the derivation of the self-energy (see below for the perturbative treatment and the self-consistent approach), it is nevertheless possible to gain some physical insight on the expected profiles of the spectral function in the different regimes of scattering. Indeed, equations~\eref{eq:spec_func_def} and \eref{eq:av_green_func} allow us to express the spectral function as
\begin{equation}
	\centering
	A(\mathcal{E},\ki)= -\frac{1}{\pi}\frac{\text{Im}[\Sigma(\mathcal{E},\ki)]}{\left( \mathcal{E} - \mathcal{E}_{k_i} - \overline{V} - \text{Re}[\Sigma(\mathcal{E},\ki)]\right)^2 + \left(\text{Im}[\Sigma(\mathcal{E},\ki)]\right)^2}.
	\label{eq:spec_func}
\end{equation}
When the scattering strength is weak, one can show that the self-energy is almost constant around the energy $\mathcal{E}_{k_i}$ and $A(\mathcal{E},\ki)$ can be approximated by a Lorentzian function (see~\fref{fig:theory_sf}(a))~\cite{Kuhn2007,Piraud2013}. As we will see in~\sref{subsec:SE_Born}, this case corresponds to the Born regime. In a more general case, the energy dependence of the self-energy must be considered and the spectral function exhibits a different profile (see~\fref{fig:theory_sf}(b)) that depends on the details of the disordered potential. 

%------------ FIGURE SF AND TAUs start ------------%
%------------ FIGURE SF AND TAUs end --------------%

When approaching infinitely large disorder strength $|\VR|$, the so-called ``classical disorder regime"~\footnote{This regime refers to the limiting case $|V_\mathrm{R} | \gg E_\sigma=\frac{\hbar^2}{m \sigma^2}$, with $E_\sigma$ the correlation energy associated to the spatial correlation $\sigma$ (see, e.g.,~\cite{Pilati2010} and references therein). For our parameters, one has $E_\sigma/h\sim 460$~Hz. }, it is again possible to predict the profile of the spectral function. Since quantum effects become negligible, the energy distribution converges towards the amplitude probability distribution $P(V)$ of the potential shifted by the kinetic energy $\mathcal{E}_{k_i}$, i.e., $A(\mathcal{E},\ki)\simeq P(\mathcal{E}-\mathcal{E}_{k_i})$ (see~\fref{fig:theory_sf}(c) for the specific case of a repulsive laser speckle disorder)~\cite{Trappe2015,Prat2016}. In that case, the FWHM $\Delta\mathcal{E}$ of the spectral function is always proportional to $|\VR|$, with a factor that depends on the specific profile of $P(V)$, yielding the  limit $\tauScl \propto 1/|\VR|$ at large disorder.  We therefore expect the scattering time $\tauS$ to be larger than the Born prediction ($\tauSB \propto 1/|\VR|^2$) when approaching the classical disorder regime, in accordance with the observations in~\fref{fig:born}.

\subsection{Perturbation theory: first order Born approximation}
\label{subsec:SE_Born}

To quantitatively estimate the self-energy, a standard method is to decompose it as an infinite sum of terms known as the Born series
\begin{equation}
	\centering
	\Sigma= \SEBornF + \SEBornS +\cdots,
	\label{eq:Born_series}
\end{equation}
each term $\Sigma_n$ involving $n+1$ occurrences of the disordered potential, as for instance  $\SEBornF= \overline{\delta V G_0 \delta V}$ or $\SEBornS= \overline{\delta V G_0 \delta V G_0 \delta V}$. Formally each term yields a contribution that corresponds to specific scattering processes, which can be illustrated using the so-called ``irreducible diagrams"~\cite{Rammer2004,Akkermans2007}. For instance the first term $\SEBornF$ only describes independent scattering events, while interference between successive scattering events is taken into account starting from the next term. Giving a detailed description of each term is beyond the scope of this paper, and we refer for instance to Ref.~\cite{Kuhn2007} for a pedagogical derivation.

As a first step, we consider in this section the first order term of the Born series~\eref{eq:Born_series}. This approximation, known as the first order Born approximation, yields $\Sigma\simeq\SEBornF= \overline{\delta V G_0 \delta V}$. This expression can be written in terms of the convolution product of the two-point correlation function $\tilde{C}(\kdis)$ and the free Green function as
\begin{equation}
	\centering
	\SEBornF(\mathcal{E},\ki)= \tilde{C}(\ki) * G_0(\mathcal{E},\ki) = \sum_{\kp} \tilde{C}(\ki-\kp) \frac{1}{\mathcal{E} - \mathcal{E}_{k'} + i0^+}.
	\label{eq:self-energy-Born}
\end{equation}
An important feature of $\SEBornF$ is that it varies slowly around the energy $\mathcal{E}_{k_i}$, such that $\SEBornF(\mathcal{E},\ki)\simeq \SEBornF(\mathcal{E}_{k_i},\ki)$ (see e.g.~\cite{Kuhn2007,Piraud2013}). As shown in~\fref{fig:theory_sf}(a), it results in that weak disorder case in a quasi Lorentzian profile 
\begin{equation}
	\centering
	A(\mathcal{E},\ki) \simeq \frac{1}{\pi}\frac{\Delta\mathcal{E}/2}{(\mathcal{E} -  \mathcal{E}_{k_i}')^2 + \Delta\mathcal{E}^2/4}.
	\label{eq:spec_func_pseudo-particle}
\end{equation} 
This function is centered around the energy $\mathcal{E}_{k_i}' \simeq \mathcal{E}_{k_i} + \overline{V} + \text{Re}[\SEBornF(\mathcal{E}_{k_i},\ki)]$ and has a FWHM $\Delta\mathcal{E}\simeq -2\text{Im}[\SEBornF(\mathcal{E}_{k_i},\ki)]$. The real part of the self-energy can then be directly interpreted as a light-shift induced by the disorder, while the imaginary part is responsible for the finite lifetime $\tauSsf=\hbar/\Delta\mathcal{E}$, with
\begin{equation}
	\centering
	\hbar/\tauSsf = -2\text{Im}[\SEBornF(\mathcal{E}_{k_i},\ki)] = 2\pi \sum_{\kp} \tilde{C}(\ki-\kp) \delta(\mathcal{E}_{k_i} - \mathcal{E}_{k'}).
\end{equation}

As announced before, we recover the prediction~\eref{eq:fermi_golden_rule} of $\tauSB$ based on the Fermi golden rule. It is expected since the first order term $\SEBornF$ is obtained by considering that the successive scattering events are independent from each other, neglecting all possible interference between them~\cite{Rammer2004,Akkermans2007}. It provides then a clear physical picture of the Ioffe-Regel like criterion $k \lS\sim 1$: when the mean free path $\lS=v\tauS$ is much larger than the de Broglie wavelength ($k\lS\gg 1$), the phase accumulated between two successive scattering events is random and the interference is washed out.

\subsection{Second order Born approximation}
\label{subsec:SE_Born2}

The comparison between our measurements and the first order Born approximation has been extensively discussed in section~\ref{sec:Born}. To go beyond, we calculate now the correction to the self-energy at the second order of perturbation $\SEBornS$, from which we deduce the correction to the width of the spectral function $-2\text{Im}[\SEBornS]$. Since it involves third-order cumulants of the potential $V(\textbf{r})$, this term vanishes for Gaussian-distributed disorder due to the symmetry of its probability distribution $P(V)$.

In contrast, it is relevant for laser speckle disorder, being of opposite signs for attractive ($\VR^3<0$) or repulsive ($\VR^3>0$) potential. The calculation in the current case of 2D laser speckle potential is detailed in~\ref{app:2nd_order}. The results are shown in~\fref{fig:born} (dotted lines), only for the three lowest disorder strengths for clarity. For attractive disorder, the correction $-2\text{Im}[\SEBornS]$ is positive, of same sign as the first order term $-2\text{Im}[\SEBornF]$, leading to a reduction of the estimated scattering time. For the lowest disorder strength $|\VR|/h=39\,\rm{Hz}$, it yields closer prediction to the numerics than the first order Born approximation $\tauSB$ (dashed lines). However, the corrected scattering time remains always smaller than $\tauSB$, while the observed $\tauS$ lies below $\tauSB$ at low $|\VR|$ but above at high $|\VR|$. Very rapidly, second order prediction deviates as well and higher order corrections must be included.

For repulsive disorder, the second order correction $-2\text{Im}[\SEBornS]$ is negative and thus of opposite sign to the first order term $-2\text{Im}[\SEBornF]$, yielding larger prediction for the scattering time. At very low disorder strength $|\VR|/h=39\,\rm{Hz}$, it is also in very good agreement with measurements of $\tauS$. However, already for relatively small disorder strength, the second order correction becomes comparable to the first order term, leading to the annulation of $\text{Im}[\Sigma]$ and a diverging prediction for $\tauS$. To restore the convergence, higher orders must be included as well.

In summary, the second order Born approximation expands the range of validity of the model only to the limited regime of low initial momenta $k_i$ and low disorder strength $|\VR|$. Qualitatively, it is expected since only first and second order terms, which scale respectively as $|\VR|^2$ and $|\VR|^3$, cannot reproduce the expected $\tauScl \propto 1/|\VR|$ scaling when approaching the strong scattering regime. Increasing the predictability range would then demand to extend the Born series to many orders. However, it is an asymptotic series that is known to diverge (see, e.g., Ref.~\cite{Kuhn2007}). There is thus an intrinsic limitation for the perturbative approach to describe $\tauS$ far beyond the weak scattering regime.

\subsection{Self-Consistent Born Approximation}
\label{subsec:SE_scba}

Rather than developing a perturbative treatment, the self-energy and thus the spectral function can be estimated using a self-consistent approach. In the first order Born approximation, the initial momentum state $\ket{\ki}$ is coupled to the free states $\ket{\kp}$ (see~\fref{fig:decay}(a)). Instead, the self-consistent Born approximation (SCBA) considers couplings to states dressed by the disorder. In order to account for the energy shift and the lifetime of those states, the self-energy is calculated by replacing the free Green function $G_0$ in~\eref{eq:self-energy-Born} with the averaged Green function $\bar{G}$, leading to the system of equations
\begin{eqnarray}
	\SEscba(\mathcal{E},\ki) = \tilde{C}(\ki) * \bar{G}_\mathrm{scba}(\mathcal{E},\ki) \label{eq:self-energy-scba}\\
	\bar{G}_\mathrm{scba}(\mathcal{E},\ki) = \left(\mathcal{E} - \mathcal{E}_{k_i} - \overline{V} - \SEscba(\mathcal{E},\ki) \right)^{-1}
	\label{eq:Green-fonction-scba}
\end{eqnarray}
that must be solved self-consistently.

%------------ FIGURE SCBA start ----------%
\begin{figure}[t]
	\includegraphics[width=0.98\textwidth]{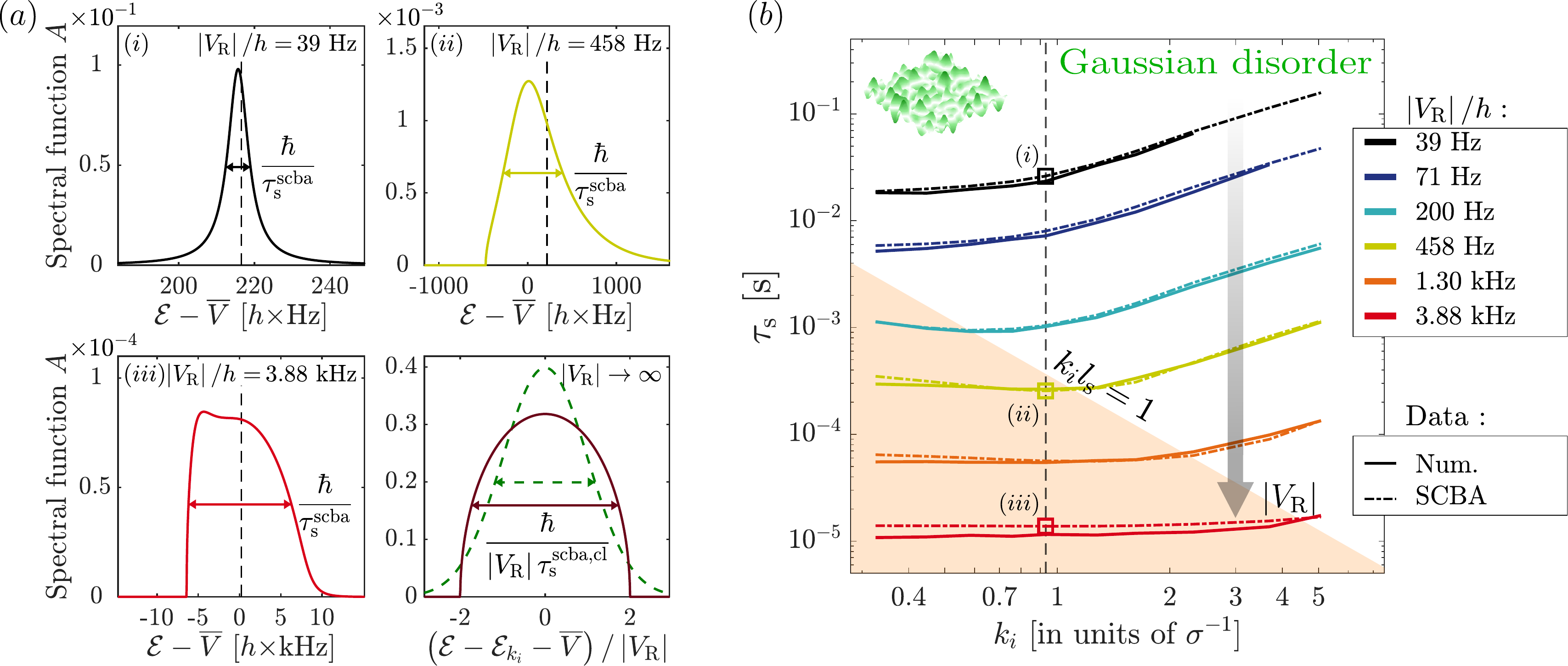}
	\caption{\label{fig:scba_gauss} \textbf{Scattering time in the self-consistent Born approximation: comparison to $\tauS$ for Gaussian-distributed disorder.} 
	(a)~Spectral functions computed with SCBA at $k_i=0.93\sigma^{-1}$ for disorder strength $|\VR|=39$ Hz (top left), $|\VR|=459$ Hz (top right) and $|\VR|=3.88$ kHz (bottom left). The vertical dashed lines indicates the kinetic energy $\mathcal{E}_{k_i}$. At infinite disorder strength (bottom right), it approaches the asymptotic function $\Ascbacl(\mathcal{E},\ki)= (2\pi \VR^2)^{-1}(4\VR^2-(\mathcal{E} - \mathcal{E}_{k_i} - \overline{V})^2)^{1/2}$ (brown solid line), that corresponds to a half circle. Besides, the infinite disorder limit for the true spectral function has a Gaussian shape (green dashed line) that reflects the amplitude probability distribution $P_g(V)$ (Gaussian-distributed disorder): it has a different profile than the limit $\Ascbacl$  but with a similar FWHM.
	(b) Comparison between numerical simulations of $\tauS$ (solid lines) and SCBA predictions $\tauSscba$ (dash-dotted lines) for Gaussian-distributed disorder. The agreement is remarkable over the whole range of parameters. The square dots refer to the parameters used to plot the spectral functions in (a).
	}
\end{figure}
%------------ FIGURE SCBA end ------------%

It is known that SCBA cannot predict the exact form of the spectral function, since, by construction, it takes only into account the two-point correlation function of the disorder $C$, regardless of the amplitude probability distribution $P(V)$~\cite{Pasek2015,Trappe2015}. Nonetheless, it is an open question whether SCBA provides a good estimate of the width of spectral function, and therefore can predict $\tauS$ better than the Born approximation. To address this question, we compute the spectral function $A_\mathrm{scba}$ in the SCBA, from which we deduce the FWHM $\Delta\mathcal{E}_\mathrm{scba}$ and its corresponding scattering time $\tauSscba=\hbar/\Delta\mathcal{E}_\mathrm{scba}$. We perform the calculation for a disorder with a Gaussian-shaped correlation function, as considered so far, to allow for comparison of $\tauSscba$ with the extracted scattering time $\tauS$.

We solve~\eref{eq:self-energy-scba} and \eref{eq:Green-fonction-scba} by iteration, up to reaching convergence (see~\ref{app:scba}). From the solution $\SEscba$, we compute the spectral function using~\eref{eq:spec_func}. \Fref{fig:scba_gauss}(a) shows examples of such spectral functions, at fixed $k_i$ and for different disorder strength $|\VR|$. At low disorder strength (top left), the spectral function almost coincides with a Lorentzian function of FWHM $\hbar/\tauSscba \simeq \hbar/\tauSB$, reproducing the expected result of the Born approximation, see~\eref{eq:spec_func_pseudo-particle}. For intermediate (top right) and strong (bottom left) disorder strength, the spectral function is broadened and its profile deviates largely from a Lorentzian distribution. In the classical limit of infinite disorder strength (bottom right), it approaches an asymptotic function $\Ascbacl$ corresponding to a semi-circle of radius $2\VR$ centered around the energy $\mathcal{E}_{k_i} + \overline{V}$~\cite{Trappe2015}. This profile sets an analytical limit for the scattering time $\tauS^\mathrm{scba,cl} = (2\sqrt{3}\VR)^{-1}$.

To benchmark the method, we compare in~\fref{fig:scba_gauss}(b) the resulting time $\tauSscba$ (dash-dotted lines) to the numerically estimated $\tauS$ in the case of Gaussian-distributed disorder (solid lines). We find that SCBA provides an excellent estimate of the scattering time over the whole range of parameters, even in the strong scattering regime $k_i\lS < 1$ (shaded area). It is particularly remarkable considering that SCBA does not reproduce spectral function in the classical limit, as illustrated in~\fref{fig:scba_gauss}(a) by comparing $\Ascbacl$ with the actual limit $P_{\mathrm{g}}$ (bottom right, dashed line). Nonetheless, the FWHMs of those two distributions are roughly similar, justifying the good agreement observed between $\tauSscba$ and $\tauS$. This is also confirmed by analytical calculation of the FWHMs, which indicates that they only differ by a close-to-unity factor $\sqrt{3/(2\text{ln}2)}\simeq 1.5$.

In this section, we have compared the scattering times, as extracted from the exponential decay of the initial momentum distribution, to the widths of the spectral functions computed by different methods. We found that, as expected, the first order Born approximation gives a good estimate for low disorder. Considering the second order term of the Born series accounts for deviations but only on a limited range of parameters. In contrast, for Gaussian disorder, the SCBA yields fair estimates for $\tauS$ while not reproducing correctly the complete spectral function. 

However, since the SCBA prediction depends only on the two-point correlation function, it cannot be sufficient to describe the behavior of $\tauS$ for arbitrary probability distribution of disordered potentials~\cite{Trappe2015}. This is especially the case for laser speckle disorders that are further discussed in next section.

\section{Elastic scattering time and real spectral functions for laser speckle disorders}
\label{sec:red_blue_sf}

We investigate here the limitations of the SCBA in describing $\tauS$ for both attractive and repulsive laser speckle potentials. A full understanding requires to explore in detail the features of the real spectral functions, which can in those cases exhibit complicated profiles. We perform such an analysis on the basis of the spectral functions measured in Ref.~\cite{Volchkov2018} in the specific case of $k_i=0$. We show that the differences reported between attractive and repulsive laser speckle disorder are at the root of the distinct behaviors that we observe on the scattering time $\tauS$.

\subsection{Limitations of SCBA prediction for laser speckle disorder}

\begin{figure}[t]
	\centering
	\includegraphics[width=0.98\textwidth]{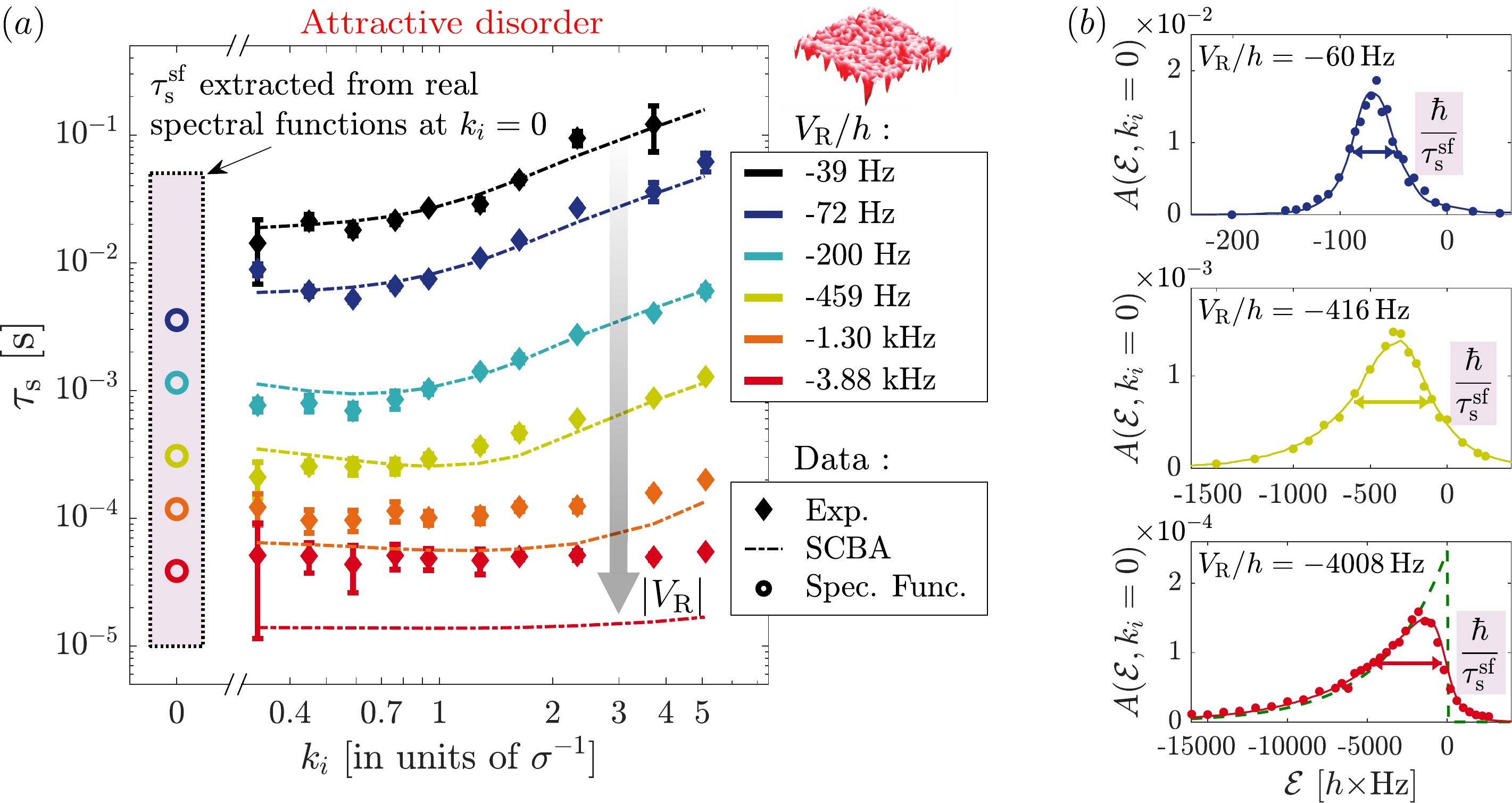}
	\caption{\label{fig:scba_red} \textbf{SCBA and real spectral functions for attractive laser speckle disorder.} 
	(a) The experimental measurements of $\tauS$ (diamonds) substantially deviate from predictions computed with SCBA (dash-dotted lines, same than in~\fref{fig:scba_gauss}). On the contrary, they are fully consistent with the values $\tauSsf$ extracted from measured spectral functions in the extreme limit $k_i=0$. For better visibility, the numerical simulations are not shown.
	(b) Experimentally measured (dots) and simulated (lines) spectral functions extracted from Ref.~\cite{Volchkov2018}, for $k_i=0$ and various disorder amplitudes $\VR$. We fit them with the convolution of an exponential and a Lorentzian distributions, and we extract $\tauSsf$ from the obtained FWHM. For the strongest disorder amplitude $\VR=-4008$~kHz, the green dashed line indicates the distribution $P_\mathrm{sp}(V)$. 
	}
\end{figure}

In~\fref{fig:scba_red}(a) we compare the SCBA prediction $\tauSscba$ to the experimental determination of $\tauS$ for attractive laser speckle disorder. Already at low initial momentum $k_i$ and low disorder strength $|\VR|$, SCBA does not perform better than the first order Born approximation. Since it only contains even powers of the fluctuations of the disorder, it does not include the second order corrections modeled by $\SEBornS$ and is thus less reliable than second order perturbation theory. 

For intermediate disorder strength $|\VR|$, SCBA yields apparently closer predictions to the experimental data than the 1st order Born approximation, but it deviates again when approaching the classical limit of strong disorder strength $|\VR|$. Although both $\tauS$ and $\tauSscba$ scale as $1/|\VR|$ in this regime, a quantitative mismatch is observed that can be attributed to the singular profile of the amplitude probability distribution $P_\mathrm{sp}(V)$ of the attractive laser speckle field. Indeed, the latter is much more peaked than $\Ascbacl$, resulting in a substantially smaller FWHM and thus larger elastic scattering time. Based on these asymptotic profiles, we calculate a ratio between $\tauScl$ and $\tauS^\mathrm{scba,cl}$ of 5, which is consistent with the value of 3.2(5) that we measure at the lowest $k_i$.

%------------ FIGURE SF RED start ----------%
%------------ FIGURE SF RED end ------------%

\begin{figure}[t]
	\includegraphics[width=0.98\textwidth]{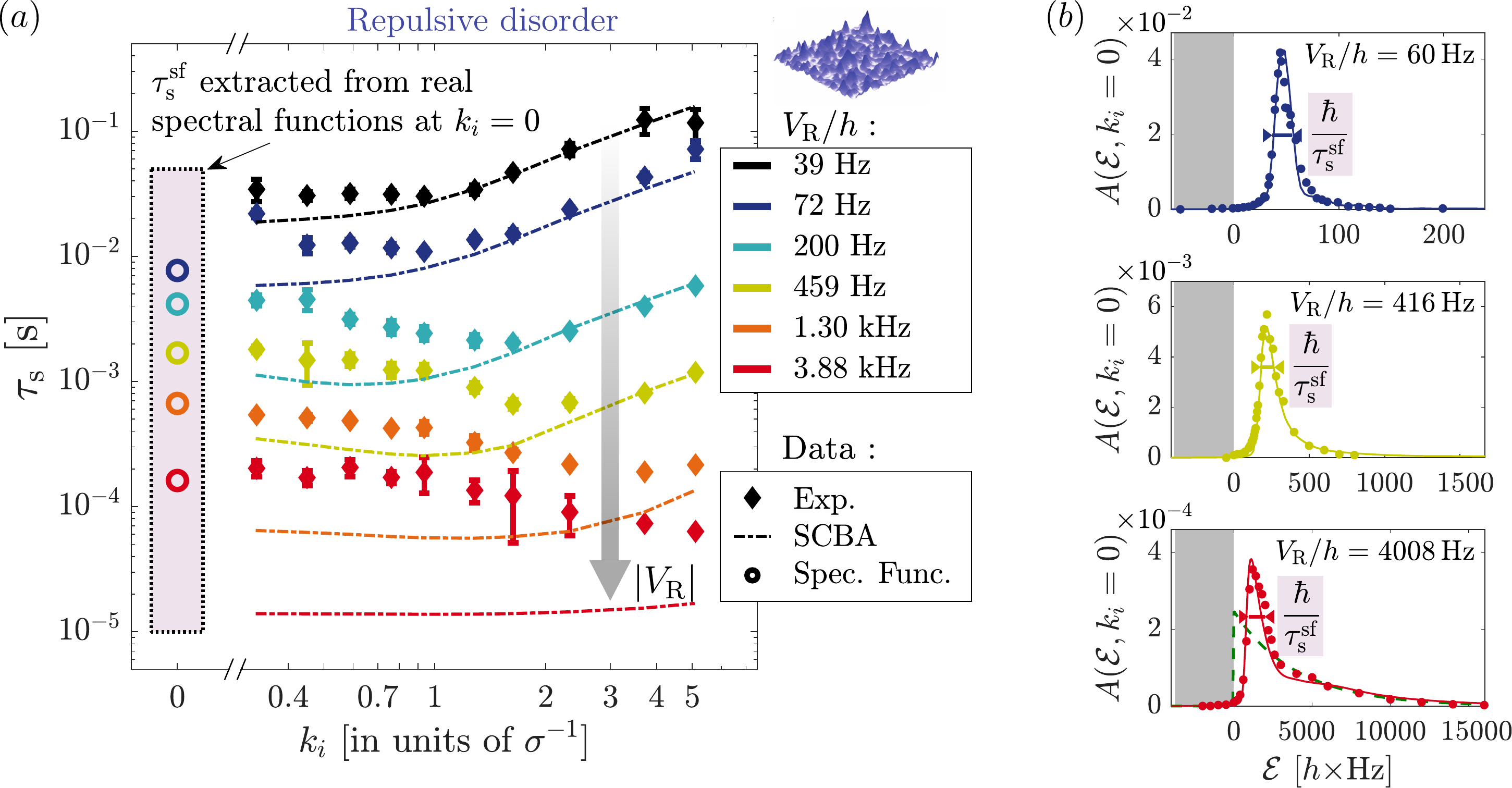}
	\caption{\label{fig:scba_blue} \textbf{SCBA and real spectral functions for repulsive laser speckle disorder.}
	(a) The experimental measurements of $\tauS$ (diamonds) strongly deviate from SCBA predictions (dash-dotted lines), while they are fully consistent with the values $\tauSsf$ extracted from measured spectral functions in the extreme limit $k_i=0$.
	(b) Experimentally measured (dots) and simulated (lines) spectral functions extracted from Ref.~\cite{Volchkov2018}, for $k_i=0$ and various disorder amplitude $\VR$. We fit them with an heuristic function, which models the bimodal structure, in order to extract $\tauSsf$. The grey area indicates the forbidden negative energies. For the strongest disorder amplitude $\VR=4008$~kHz, the green dashed line indicates the distribution $P_\mathrm{sp}(V)$.
	}
\end{figure}

The case of repulsive laser speckle disorder is shown in~\fref{fig:scba_blue}(a). At low initial momentum $k_i$, low disorder strength $|\VR|$, deviations of similar magnitude compared to an attractive laser speckle are observed, originating from the same absence of odd order correction terms. For increasing disorder strength $|\VR|$, however, deviations become much more pronounced, reaching more than 1 order of magnitude at the largest disorder strength. This cannot be simply justified by the profile of the amplitude probability distribution and therefore requires deeper analysis of the profile of the spectral functions.

%------------ FIGURE SF BLUE start ----------%
%------------ FIGURE SF BLUE end -----------%

\subsection{Comparison with measured spectral functions}

To further investigate the different behaviors of $\tauS$, a comparison to the real spectral functions is needed. Experimentally, spectral functions of matter-waves in laser speckle disorder have been measured in the specific case $k_i=0$ and for a large set of disorder strength $|\VR|/h$ ranging from 60 Hz to 4 kHz~\cite{Volchkov2018}. Three examples are shown in~\fref{fig:scba_red}(b) in the case of attractive speckle potential. At weak disorder strength $|\VR|/h=60$~Hz (top panel), the spectral function exhibits an approximately Lorentzian profile, consistent with the Born interpretation. The profile changes at intermediate disorder strength (central panel) to approach for strong disorder the classical limit $P_\mathrm{sp}(V)$ (green dashed line in the bottom panel), although deviations due to quantum corrections still persist around $\mathcal{E} \sim 0$~\cite{Volchkov2018}. 
For those measured spectral functions, we extract the FWHM from a fit and we deduce the elastic scattering time $\tauSsf$ in the limit $k_i=0$ as a function of $|\VR|$ (\ref{app:tauS_sf}). The results, plotted in~\fref{fig:scba_red}(a) (circle dots) for the same values of $|\VR|$ than considered so far. They are in good agreement with the low momentum limit of $\tauS$, especially at strong disorder. This shows that our measurements of $\tauS$ are fully consistent with the specific profiles of the spectral functions for the attractive laser speckle case.

For repulsive laser speckle, the spectral functions at $k_i=0$ are plotted in~\fref{fig:scba_blue}(b). Since negative energies are strictly forbidden in repulsive potential (depicted by the grey area), the profiles are intrinsically different in comparison to the previous case. At weak disorder strength, spectral functions still follow Lorentzian-like profile typical of the Born regime. In the strong disorder regime, they exhibit a narrow resonance peak on top of the broad distribution that would have been expected from the classical limit $P_\mathrm{sp}$ (green dashed line). The presence of this peak is related to an accumulation of bound states around the averaged ground state harmonic oscillator energy~\cite{Trappe2015,Prat2016,Volchkov2018}. 

As a consequence of this double structure, the time evolution is expected to show two different timescales: a short one associated to the broad part of the spectral function and a long one associated to the narrow peak. Experimentally, the measured time evolution is dominated by the slowest decay. The characteristic time we have extracted when measuring $\tauS$ is thus related to the long timescale, and should be compared to the FWHM of the narrow peak. 
To perform the comparison, we fit the spectral functions by an heuristic function accounting for the bimodal structure (see~\ref{app:tauS_sf}), and we extract $\tauSsf$ from the FWHM of the peaked function. As shown in~\fref{fig:scba_blue}(a), it agrees once again very well with the low momentum limit of $\tauS$.

In conclusion, analysing the profiles of the real spectral functions allows us to interpret the observed differences in the scattering time $\tauS$ between attractive and repulsive laser speckle disorder. The striking agreement between time domain and energy domain measurements validates our method to extract $\tauS$ in a broad range of scattering regimes. It also highlights the key issue here, which is to find theoretical models that reproduce the specific features of the spectral functions~\cite{Trappe2015,Prat2016}.

\section{Summary and outlook}
\label{sec:conclusion}

We have investigated in this paper the behavior of the elastic scattering time $\tauS$ of ultracold atoms in disordered potentials in the strong scattering regime. A first important result is the remarkable agreement between the observed behavior of $\tauS$ and predictions based on the self consistent Born approximation (SCBA), in the case of Gaussian-distributed disordered potentials. However, this method, which inherently doesn't take into account the specific form of the disorder amplitude distribution, is not accurate for laser speckle disorder. Instead, we have shown that the calculation of the second order term in the Born series, which is sensitive to the distribution skewness, is able to explain the differences reported between the attractive and the repulsive speckle disorders when entering the strong scattering regime. The validity of this pertubative approach is nevertheless limited to a narrow range of parameters. As a second main result, we show that one has then to rely on the real shape of the spectral functions, as measured in~\cite{Volchkov2018}, in order to interpret our data.

Altogether, our study clarifies the validity range of common theoretical methods to predict the elastic scattering time of matter waves in disordered potential.  It highlights the need for developing adequate formalisms in order to cope with the full statistics of the disorder, especially in the case of laser speckle disorder that are commonly used with ultracold atoms. Beside semiclassical approaches, dedicated to the asymptotic classical regime~\cite{Trappe2015,Prat2016}, or the coherent potential approximation~\cite{Maier2005,Zimmermann2009,Pasek2015}, a very interesting follow up would  be to confront our measurements to the recent theoretical framework of the hidden landscape~\cite{Filoche2012,Arnold2016}. Such development of quantitative predictions is essential for the understanding of complex transport phenomena, such as the Anderson localization where discrepancies remain between experiments, numerics and available theories (see e.g.~\cite{pasek2016anderson}).

%--------------Ackowledgment----------------%

\ack
We would like to thank D. Delande, T. Giamarchi, G. Montambaux, C. M\"uller and S. Skipetrov  for fruitful discussions. This work has been supported by ERC (Advanced Grant ``Quantatop''), Institut Universitaire de France, the French Ministry of Research and Technology (ANRT, through a DGA grant for J.~R. and CIFRE/DGA grant for V.~D.), the EU-H2020 research and innovation program (Grant No. 641122-QUIC and Marie Sk\l odowska-Curie Grant No. 655933), Region Ile-de-France in the framework of DIM SIRTEQ and the Simons foundation (Grant No. 563916: Localization of waves).

%------------------ APPENDIX -----------------------%
\appendix

\section{Second order Born approximation for laser speckle disorder} \label{app:2nd_order}

We show here how to calculate the second order correction $\SEBornS=\overline{\delta V G_0 \delta V G_0 \delta V}$ of the Born expansion. It involves the three-point correlation function $C_3(\Delta \mathbf{r},\Delta  \mathbf{ r'})=\overline{\delta V(\mathbf{r})\delta V(\mathbf{r} +\Delta \mathbf{r})\delta V(\mathbf{r}+\Delta \mathbf{r'})}$, which is strictly null for Gaussian-distributed disorder and equals to

\begin{equation}
	C_3(\Delta \mathbf{r},\Delta \mathbf{ r'})= 2\VR^3 \e^{-[\Delta \mathbf{r}^2 + \Delta \mathbf{r'}^2 + (\Delta \mathbf{ r} -\Delta \mathbf{ r'})^2]/(2\sigma^2)}
	\label{eq:3pts_correlation-fct}
\end{equation}
for a 2D laser speckle disordered potential. 

The operator $\SEBornS$ is diagonal in the $\ket{\bfk}$ basis, with matrix elements $\SEBornS(\mathcal{E},\ki)=\langle \ki | \SEBornS | \ki \rangle$ given by
\begin{equation}
	\SEBornS(\mathcal{E},\ki) = \sum_{\kp,\ks} \tilde{C}_3(\ki-\kp,\ki-\ks) G_0(\mathcal{E}, \mathcal{E}_{k'}) G_0(\mathcal{E}, \mathcal{E}_{k''}),
	\label{eq:Sigma2}
\end{equation}
where $\tilde{C}_3(\kdis,\kdis')$ is the Fourier transform of the three-point correlation function~\eref{eq:3pts_correlation-fct}. 
When taking the imaginary part of~\eref{eq:Sigma2}, we obtain

\begin{eqnarray}
\nonumber \fl \text{Im}[\SEBornS(\mathcal{E},\ki)] = -\frac{8\pi^3\VR^3}{3} \sum_{\kp,\ks} & \e^{-[(\ki-\kp)^2 + (\kp-\ks)^2 + (\ks-\ki)^2]\sigma^2/6} \\
& \left[ \delta(\mathcal{E}-\mathcal{E}_{k'}) \text{p.v.}(\mathcal{E}-\mathcal{E}_{k''}) + \text{p.v.}(\mathcal{E}-\mathcal{E}_{k'}) \delta(\mathcal{E}-\mathcal{E}_{k''}) \right],
\label{eq:ImSigma2}
\end{eqnarray}
where p.v.~refers to the Cauchy principal value. The two terms revealed by~\eref{eq:ImSigma2} are related to the two possible third-order processes, corresponding either to a single scattering event for the wavefunction and two scattering events for the conjugated wavefunction, or to the other way around. 

The second order Born correction is finally obtained by numerically calculating~\eref{eq:ImSigma2} at the energy $\mathcal{E}=\mathcal{E}_{k_i}$.

\section{Calculation of the self-energy in the SCBA} \label{app:scba}

We present in this section the main stages in the calculation of the self-energy $\SEscba(\mathcal{E},\ki)$ in the self-consistent Born approximation. The procedure is detailed for a given set of disorder strength $|\VR|$ and initial momentum $\ki$, the overall process being repeated for all the sets of parameters we have explored.

At first we define an energy range $[\mathcal{E}_\mathrm{min},\mathcal{E}_\mathrm{max}]$ relevant for the calculation of the spectral function. It is chosen to be centered on the kinetic energy $\mathcal{E}_{k_i}$, with a width large enough to ensure that the spectral function area is close to unity. For each energy $\mathcal{E}$ of this interval, the self-consistent equations are then solved. To do so, we calculate in parallel the self-energy $\SEscba(\mathcal{E},\bfk)$ for all the momenta $\bfk$ whose kinetic energies are contained in the range $[\mathcal{E}_\mathrm{min},\mathcal{E}_\mathrm{max}]$. We proceed by iteration, initializing the solution with the first order Born solution $\SEscba^{(0)}=\SEBornF$. Using~\eref{eq:self-energy-scba} and~\eref{eq:Green-fonction-scba}, the self-energy $\SEscba^{(n+1)}$ after $n+1$ iteration steps is given by
\begin{equation}
	\SEscba^{(n+1)}(\mathcal{E},\bfk) = \int{ \frac{\rmd^2\kp}{4\pi^2} \tilde{C}(\bfk - \kp) \frac{1}{ \mathcal{E} - \mathcal{E}_{k'} - \SEscba^{(n)}(\mathcal{E},\kp)} }.
\end{equation}
The iteration loop is pursued until the convergence criterion
\begin{equation}
	\int{ \frac{\rmd^2\bfk}{4\pi^2} \left|\frac{\SEscba^{(n+1)}(\mathcal{E},\bfk) - \SEscba^{(n)}(\mathcal{E},\bfk) }{ \SEscba^{(n+1)}(\mathcal{E},\bfk) }\right| } <10^{-3}
\end{equation}
is reached.

Once the self-energy is known for each energy $\mathcal{E}$, the spectral function can be computed using~\eref{eq:spec_func}. Its normalization is verified to make sure the energy interval was correctly chosen.

\section{Extracting the widths of real spectral functions} \label{app:tauS_sf}

We present here the procedure to extract the widths $\Delta \mathcal{E}$ of the spectral functions experimentally measured in Ref.~\cite{Volchkov2018}, in order to estimate the scattering time $\tauSsf$. Since the spectral functions are noticeably different between attractive and repulsive laser speckle potential, we distinguish the two cases on the following.

For attractive laser speckle disorder, we use as fit function the convolution of a Lorentzian distribution $L$ with an exponential distribution $R$:
\begin{equation}
	{A}_\mathrm{fit}^\mathrm{att}(\mathcal{E}) = {L}(\mathcal{E},\mathcal{E}_c,\delta\mathcal{E}) * {R}(\mathcal{E},\delta\mathcal{E}').
	\label{eq:Afit_att}
\end{equation}
The Lorentzian distribution
\begin{equation}
	L(\mathcal{E},\mathcal{E}_c,\delta\mathcal{E}) = \frac{1}{\pi} \frac{\delta \mathcal{E}/2}{(\mathcal{E}-\mathcal{E}_c)^2 + \delta \mathcal{E}^2/4}
\end{equation}
has a central energy $\mathcal{E}_c$ and a full-width at half-maximum $\delta\mathcal{E}$. The exponential distribution is defined as
\begin{equation}
	R(\mathcal{E},\delta\mathcal{E}') = \frac{1}{\left|\delta\mathcal{E}'\right|} \rme^{-\mathcal{E}/\delta\mathcal{E}'}\Theta{(\mathcal{E}/\delta\mathcal{E}')},
\end{equation}
such that it converges towards the amplitude distribution of the attractive disorder $P_\mathrm{sp}$ when its width $\delta\mathcal{E}'$ approaches $\VR<0$.

At low disorder strength $|\VR|$, the width $\delta\mathcal{E}'$ goes to 0 and ${A}_\mathrm{fit}^\mathrm{att}$ approaches a Lorentzian profile, as experimentally observed (see top panel in~\fref{fig:scba_red}(b)). At high disorder strength, $\delta\mathcal{E}'$ goes to $\VR$ for $R(\mathcal{E},\delta\mathcal{E}')$ to converge towards the classical limit $P_\mathrm{sp}$ (green dashed line in~\ref{fig:scba_red}(b)). The convolution with $L$ guarantees that the fit function is smoothed around $\mathcal{E}=0$, with $\delta\mathcal{E}$ corresponding almost to the energy region where quantum effects are relevant. In between these two extreme cases, ${A}_\mathrm{fit}^\mathrm{att}$ reproduces well all the profiles of the measured spectral functions.

For each disorder strength $|\VR|$ for which the spectral function has been measured, we extract the FWHM $\Delta\mathcal{E}$ of $A_\mathrm{fit}^\mathrm{att}$ and we deduce the scattering time $\tauSsf = \hbar/\Delta\mathcal{E}$. The value of $\tauSsf$ for any disorder strength is then deduced by interpolation.

%------------- BLUE ----------------
The case of repulsive disorder is more complicated. The profiles of the spectral functions are bimodal, made of a narrow peak at low energy on top of a broad energy distribution. 
We fit those profiles by the sum of a convoluted distribution $L*R$ accounting for the broad part of the spectrum, and a Gaussian distribution $G$ accounting for the narrow peak. It yields for the fit function
\begin{equation}
	{A}_\mathrm{fit}^\mathrm{rep}(\mathcal{E}) = \alpha {L}(\mathcal{E},\mathcal{E}_c,\delta\mathcal{E}) * {R}(\mathcal{E},\delta\mathcal{E}') + (1-\alpha) {G}(\mathcal{E},\mathcal{E}_c,\delta\mathcal{E}),
	\label{eq:Afit_rep}
\end{equation}
with $\alpha$ the relative weight of the first contribution. The Gaussian distribution is defined as
\begin{equation}
	G(\mathcal{E},\mathcal{E}_c,\delta\mathcal{E}) = \frac{1}{\sqrt{\pi}\delta\mathcal{E}} \rme^{-\left(\frac{\mathcal{E}-\mathcal{E}_c}{\delta\mathcal{E}}\right)^2}.
\end{equation}
where the central energy $\mathcal{E}_c$ and the full width $\delta\mathcal{E}$ are chosen to be the same as the ones of $L$. It results in total in only one more free parameter -- $\alpha$ -- compared to the case of attractive disorder. 
Since the bimodal profile are more pronounced in the numerical simulations, $\alpha$ is extracted on the numerical data and kept fixed when fitting the experimental data. 

At low $|\VR|$, the spectral function is dominated by a narrow peak that is fitted by $G$, while $L*R$ accounts for the small, broad background. When increasing $|\VR|$, the peak amplitude decreases and the spectral function approaches closer to its classical limit. At high $|\VR|$, the broad part of the spectrum resembles the one of the attractive case, whose features are captured by $L*R$, while the narrow peak resulting from the accumulation of bound levels is fitted by $G$. 
Overall, the fit function~\eref{eq:Afit_rep} shows remarkable agreement with the measured spectral functions at any disorder strength. Since we are experimentally sensitive to the longer timescale when measuring $\tauS$, the relevant energy scale $\Delta\mathcal{E}$ is given by the width of the narrower structure $\delta\mathcal{E}$.

Similarly to the attractive laser speckle case, the value of $\tauSsf$ for any disorder strength is extracted by interpolation of those measurements.

%------------------ BIBLIOGRAPHY -----------------------%
\section*{Bibliography}

\bibliographystyle{h-physrev}
%\bibliographystyle{unsrtnat}
%\bibliography{TauS_PRLpaper_corr} % TauS_NJP_VJ

\end{document}